\documentclass[fleqn,usenatbib]{mnras}

\usepackage{newtxtext,newtxmath}
\usepackage[T1]{fontenc}
\DeclareRobustCommand{\VAN}[3]{#2}
\let\VANthebibliography\thebibliography
\def\thebibliography{\DeclareRobustCommand{\VAN}[3]{##3}\VANthebibliography}
\usepackage{graphicx}	
\usepackage{amsmath}	

\title[The Phase-Space of UDGs in the Hydra~I Cluster]{Ultra diffuse galaxies in the Hydra~I cluster from the LEWIS Project:  Phase-Space distribution and globular cluster richness}

\author[D. A. Forbes et al.]{Duncan A. Forbes, $^{1}$\thanks{E-mail: dforbes@swin.edu.au}
Jonah Gannon$^{1}$, Enrichetta Iodice$^2$, Michael Hilker$^5$,Goran Doll$^{2,3}$, Chiara Buttitta$^2$, 
\newauthor
Antonio La Marca$^{6,7}$, 
Magda Arnaboldi$^5$, 
Michele Cantiello$^{4}$,
G. D'Ago$^8$,  
Jesus Falcon Barroso$^{15,16}$,
\newauthor
Laura Greggio$^{9}$, 
Marco Gullieuszik$^{9}$, 
Johanna Hartke$^{12,13}$,
Steffen Mieske$^{10}$,
Marco Mirabile$^{4,14}$,
\newauthor
Roberto Rampazzo$^{9}$,
Marina Rejkuba$^5$, 
Marilena Spavone$^2$, 
Chiara Spiniello$^{11}$, 
Giulio Capasso$^2$ 
\\
$^{1}$ Centre for Astrophysics \& Supercomputing, Swinburne University, Hawthorn, VIC 3122, Australia\\
$^{2}$ INAF - Astronomical Observatory of Capodimonte, Salita Moiariello 16, I-80131, Naples, Italy\\
$^{3}$ University of Naples “Federico II”, C.U. Monte Sant’Angelo, Via Cinthia, 80126, Naples, Italy\\
$^{4}$ INAF - Astronomical Observatory of Abruzzo, Via Maggini, 64100, Teramo, Italy\\
$^{5}$ European Southern Observatory, Karl-Schwarzschild-Strasse 2, 85748 Garching bei Muenchen, Germany\\
$^{6}$ SRON Netherlands Institute for Space Research, Landleven 12, 9747 AD Groningen, The Netherlands\\
$^7$ Kapteyn Astronomical Institute, University of Groningen, Postbus 800, 9700 AV Groningen, The Netherlands\\
$^8$ Institute of Astronomy, University of Cambridge, Madingley Road, Cambridge, CB3 0HA\\
$^{9}$ Osservatorio Astronomico di Padova, Via dell'Osservatorio
8, 36012 Asiago (VI), Italy\\
$^{10}$ European Southern Observatory, Alonso de Cordova 3107, 7630355 Vitacura, Santiago, Chile\\
$^{11}$ Sub-department of Astrophysics, University of Oxford, 
Denys Wilkinson Building, Keble Road, Oxford OX1 3RH, United Kingdom  \\ 
$^{12}$ Finnish Centre for Astronomy with ESO (FINCA), FI-20014 University of Turku, Finland\\
$^{13}$ Tuorla Observatory, Department of Physics and Astronomy, FI-20014 University of Turku, Finland\\
$^{14}$ Gran Sasso Science Institute, L’Aquila, Italy\\
$^{15}$ INAF - Astronomical Observatory of Padova, Vicolo dell'Osservatorio 5, I-35122 Padova, Italy\\
$^{15}$ Instituto de Astrof\'isica de Canarias, V\'ia L\'actea s/n, E-38205 La Laguna, Tenerife, Spain\\
$^{16}$ Departamento de Astrof\'isica, Universidad de La Laguna, E-38200 La Laguna, Tenerife, Spain
}

\date{Accepted XXX. Received YYY; in original form ZZZ}

\pubyear{2023}

\begin{document}
\label{firstpage}
\pagerange{\pageref{firstpage}--\pageref{lastpage}}
\maketitle

\begin{abstract}
Although ultra diffuse galaxies (UDGs) are found in large numbers in clusters of galaxies, the role of the cluster environment in shaping their low surface brightness and large sizes is still uncertain. Here we examine a sample of UDGs in the Hydra~I cluster (D = 51 Mpc) with new radial velocities obtained as part of the LEWIS (Looking into the faintest with MUSE) project using VLT/MUSE data. Using a phase-space, or infall diagnostic, diagram we compare the UDGs to other known galaxies in the Hydra~I cluster and to UDGs in other clusters. The UDGs, along with the bulk of regular Hydra~I galaxies,  have low relative velocities and are located near the cluster core, and thus consistent with very early infall into the cluster. Combining with literature data, we do not find the expected trend of GC-rich UDGs associated with earlier infall times. This result suggests that quenching mechanisms other than cluster infall should be further considered, e.g. quenching by strong feedback or in cosmic sheets and filaments.
Tidal stripping of GCs in the cluster environment also warrants further modelling.

\end{abstract}


\section{Introduction} \label{sec:intro}

The possible formation pathways of ultra-diffuse galaxies (UDGs) have been a subject of an ongoing vigorous debate since 2015, when a population of these extremely diffuse galaxies was identified in the Coma cluster using the Dragonfly Telephoto Array \citep{2015ApJ...798L..45V}. 
Existing in all environments, they are most common in clusters with several hundred found in the Coma cluster \citep{2016ApJS..225...11Y, 2018MNRAS.479.3308A}.
This significant contribution to our ‘census of galaxies’ has prompted numerous simulation studies and
accompanying predictions (see 
\cite{2020MNRAS.494.1848S} and references therein). These  simulations can be broadly placed in two categories; internal processes
(e.g. episodic supernova feedback) or external (e.g. tidal effects in a dense environment). Some combination of
both processes may be operating along with past galaxy infall (and subsequent quenching) into clusters.

UDGs have low surface brightnesses (they are defined to have central values in the g band of $\mu$ $>$ 24 mag. per sq. arcsec) so that spectroscopic
studies of them push even 8–10m class telescopes, with efficient low surface brightness instruments, 
such as KWCI on Keck or MUSE on VLT, to their limits. While strictly speaking dwarf galaxies with
M$_{\ast}$ $< 10^9$ M$_{\odot}$, UDGs are unlike classical dwarfs as they have extreme sizes with effective radii R$_e$
$>$ 1.5 kpc (i.e. comparable to the disk of the Milky Way with R$_e$ $\sim$3.5 kpc). They also reveal another unexplained feature, with some 
hosting up to ten times more globular clusters (GCs) than classical dwarf galaxies of the same luminosity \citep{2020MNRAS.492.4874F}. 
Their
very existence in clusters and their generally old stellar populations suggests that some may be protected within an overly massive dark
matter halo. The latter is supported by the correlation between GC numbers and host galaxy halo mass for normal galaxies, e.g. 
\cite{2020AJ....159...56B}. 

In the standard picture of dwarf galaxy evolution
\citep{2016MNRAS.455.2323M}, dwarfs that fell into clusters at early times will have experienced intense star formation, prior to, or at the start of, infall (which is also expected to give rise to a high fraction of stars in bound star clusters). This is  followed by quenching of any further star formation as the infall proceeds. Both of these effects would lead to a high fraction of GCs relative to their field stars 
\citep{2016MNRAS.455.2323M, 2020MNRAS.493.5357R}.  
Indeed trends of GC richness and [$\alpha$/Fe] ratios
with clustercentric radius provide some  observational support for this interpretation \citep{2008ApJ...681..197P, 2016ApJ...818..179L, 2023MNRAS.522..130R}. 
This early-infall, or biasing, has been invoked for UDGs by \cite{2021MNRAS.502..398C} 
who include
cluster tidal effects within the IllustrisTNG simulation and simplified GC formation physics. Similar to
classical dwarfs, they predict that early-infall UDGs should be rich in GCs. Based on a semi-empirical model, \cite{2022MNRAS.510.3356T} also predict that galaxies near the cluster core form more GCs. 

Using phase-space, or infall diagnostic, diagrams of the type proposed by \cite{2017ApJ...843..128R} 
one can investigate whether GC richness depends on UDG cluster infall time. No trend between GC richness and very early infall times might suggest that GC formation and quenching occurred before cluster infall. While low mass galaxies typically quench at late times, there is a considerable range in quenching times with some low mass galaxies quenching at z $\sim$ 2 or  
10.5 Gyr ago \citep{2020MNRAS.499.4748M}. 
Quenching at early times via stellar feedback \citep{2013MNRAS.428..129S} may be one possibility. This early quenching applied to UDGs has been described by 
\cite{2022ApJ...927L..28D}.
Another possibility may be quenching via the  interaction with cosmic sheets or filaments \citep{2023MNRAS.520.2692P}.
A first attempt at this sort of infall analysis applied to GCs was presented in \citet{2022MNRAS.510..946G} 
for several UDGs in the Coma and Perseus clusters. No clear signal was found but the sample was small with just over a dozen UDGs and with a bias towards GC-rich UDGs.

In this Letter we examine the infall diagnostic diagram for a new sample of UDGs in the Hydra~I cluster (A1060; D = 51 $\pm$ 4 Mpc). The Hydra~I cluster appears to be fairly dynamically relaxed \citep{2011A&A...528A..24V} but also reveals hints of substructures \citep{2021MNRAS.500.1323L, 2022A&A...659A..92L}, an infalling group of galaxies \citep{2012A&A...545A..37A}, 
and evidence for ram pressure stripping 
\citep{2021ApJ...915...70W, 2021A&A...652L..11I}.
The observed UDGs are located near the cluster core and the northern subgroup, with all lying within the 0.3 virial radii (R$_{200}$) of the Hydra~I cluster centre. Each was observed using MUSE on the VLT as part of the ongoing LEWIS (Looking into the faintEst WIth MUSE) project. Details of the project, including galaxy radial velocities, positions, GC counts etc, are given in Paper I by Iodice et al. (2023, in press). 
The GC counts are based on deep, optical multi-filter imaging with the VST as part of the VEGAS project
\citep{2020A&A...642A..48I} and will be updated after the full analysis of the MUSE data. 
Here we explore the distribution of UDGs in phase space and investigate whether they reveal any trend in this space with their GC richness. We also include similar data for UDGs in other nearby clusters. For the Hydra~I cluster we adopt the same parameters as used by \cite{2022A&A...665A.105L}, i.e. cz = 3683 $\pm$ 46 km s$^{-1}$, $\sigma$ = 724 $\pm$ 31 km s$^{-1}$, and virial parameters R$_{200}$ = 1.6 Mpc, M$_{200}$ = 2 $\times$ 10$^{14}$ h$^{-1}$ M$_{\odot}$ and take its centre as NGC 3311 (RA = 159.17842, Dec = --27.528339). These values are similar to those found by \cite{2021MNRAS.500.1323L} who recently studied Hydra~I galaxies out to the virial radius. 

\begin{figure*}
	\includegraphics[width=\linewidth]{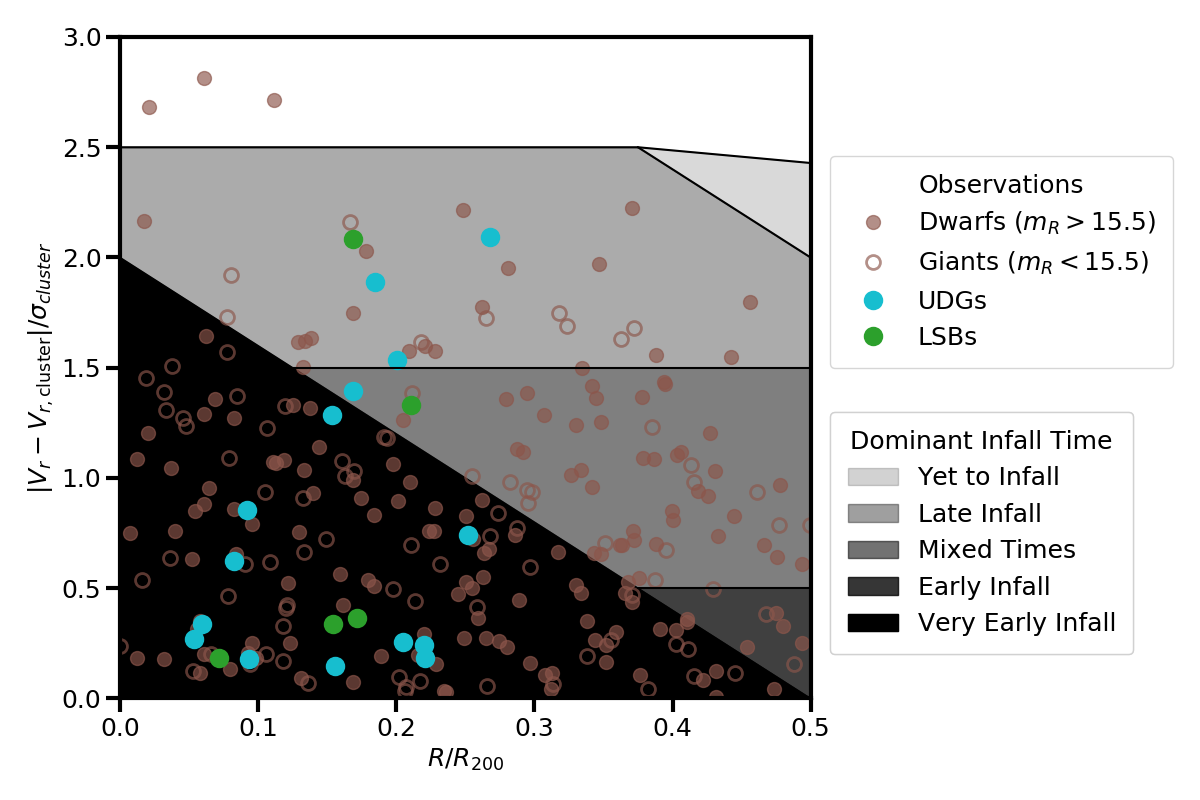}
	\caption{Infall diagnostic diagram for non-UDG (giants, dwarfs and LSB galaxies) and UDGs in the Hydra~I cluster. The diagram shows the relative line-of-sight velocity of each galaxy normalised by the cluster velocity dispersion against the projected radius normalised by the virial radius. Regions of the diagram are shaded  according to their infall times from the cosmological simulations of 
 Rhee et al. (2017) 
 as indicated in the legend. The plot shows that most UDGs and non-UDG galaxies of the the Hydra I cluster lie within the very early infall zone -- the simulations of 
 indicating that around half of the galaxies in this zone were part of the cluster at least 6.45 Gyr ago. 
 }
\end{figure*}

\section{Infall Diagnostic Diagram for Hydra~I Cluster Galaxies} 


\citet{2017ApJ...843..128R}
carried out cosmological simulations of several galaxy clusters and examined the resulting distribution of galaxies in phase-space (i.e. velocity of the galaxy relative to the mean cluster velocity normalised by the cluster velocity dispersion versus the galaxy's clustercentric radius normalised by the cluster virial radius). Based on the infall time of galaxies, they divided this diagram into several infall zones, ranging from those that fell into the cluster at very early times, to those that are yet to fall in. Thus the location of galaxies in this diagram provides an `infall diagnostic' which is statistical in nature and additional scatter is introduced when using 2D projected radii (as is the case for observational data). For example, the `very early infall' (or ancient infaller) zone in the simulation is occupied by a slight majority (52\%) of galaxies that have resided in the cluster for more than 6.45 Gyr. Projection effects mean that the true clustercentric radius for some galaxies is larger in 3D than observed in 2D. For 
most galaxies this effect should be 
less a factor of two from the projected one.

In Fig. 1 we show such an infall diagnostic diagram for all galaxies in the Hydra~I cluster out to half the virial radius. This includes 
giant and dwarf galaxies from the study of 
\cite{2003ApJ...591..764C} 
plus the addition of UDGs and 3 low surface brightness (LSB) galaxies that have UDG-like sizes but are slightly brighter from Iodice et al. (2023, in press). 
We find that the bulk of the non-UDG Hydra~I galaxies 
are located within the `very early infall' zone. The simulation of \cite{2017ApJ...843..128R}  
predicts that just over half of these would have been part of the cluster for at least 6.45 Gyr. There are also galaxies located in later infall zones 
and three galaxies that may lie outside of the cluster with large relative velocities -- these could be backsplash galaxies (having passed through the cluster) or simply galaxies that are yet to fall into the cluster.

If we examine giant and classical dwarf galaxies separately (divided at M$_R$ = --18 or m$_R$ = 15.5) there is no clear difference between them in terms of their infall properties. 
Compared to the UDGs they appear to scatter to higher relative velocities on average. 
 A more quantitative measure of the differences in their infall properties can be obtained from the product of their relative velocity from the cluster mean and their radial position: $\Delta$V/$\sigma$ x R/R$_{200}$. 
Restricting to 0.3R/R$_{200}$,  as probed by the imaging, 
we find 
mean values (and error on the mean) of $\Delta$V/$\sigma$ x R/R$_{200}$ = 0.83 ($\pm$ 0.07) $\times$ 0.15 ($\pm$ 0.01) = 0.12 ($\pm$ 0.02)  
for giant galaxies 
and 0.88 ($\pm$ 0.06) $\times$ 0.16 ($\pm$ 0.01) = 0.14 ($\pm$ 0.02) for classical dwarfs. 
For the UDGs the mean value is $\Delta$V/$\sigma$ x R/R$_{200}$ = 0.80 ($\pm$ 0.17) $\times$ 0.16 ($\pm$ 0.02) = 0.13 ($\pm$ 0.04). This indicates 
that UDGs are similarly concentrated in phase-space to the other cluster galaxies. Also, 
while UDGs have a similar distribution in clustercentric radius, their velocities are closer to the cluster mean than either giants or classical dwarfs. 
We note that  \cite{2021MNRAS.500.1323L} also found passive early-type galaxies to be  concentrated in the cluster core.The LSB galaxies in Fig. 1 are found in a range of infall zones, from early to late infall. 

As might be expected from their inner cluster position, our UDGs were among the earliest inhabitants of the cluster, infalling at least 6.45 Gyr ago according to simulations of \cite{2017ApJ...843..128R}. 
They would be expected to have star formation (SF) histories that indicate early quenching. 
A preliminary analysis by Iodice et al. (2023, submitted) for one UDG (UDG11) indicates an old age of 
$\sim$10 Gyr, 
suggestive of early quenching. 
Future analysis will also include the [$\alpha$/Fe] ratios which appears to be a sensitive indicators of SF histories for low mass galaxies (see Ferre-Mateu et al. 2023, submitted for results on UDGs in other clusters and 
\cite{2023MNRAS.522..130R}, 
for dwarf galaxies in the Fornax cluster). 
We note that the study of \cite{2021MNRAS.500.1323L} found 88\% of Hydra~I galaxies (with log M$_{\ast}$ $>$ 8.5) to be quenched, i.e. no sign of ongoing star formation. 

\begin{figure*}
	\includegraphics[width=\linewidth]{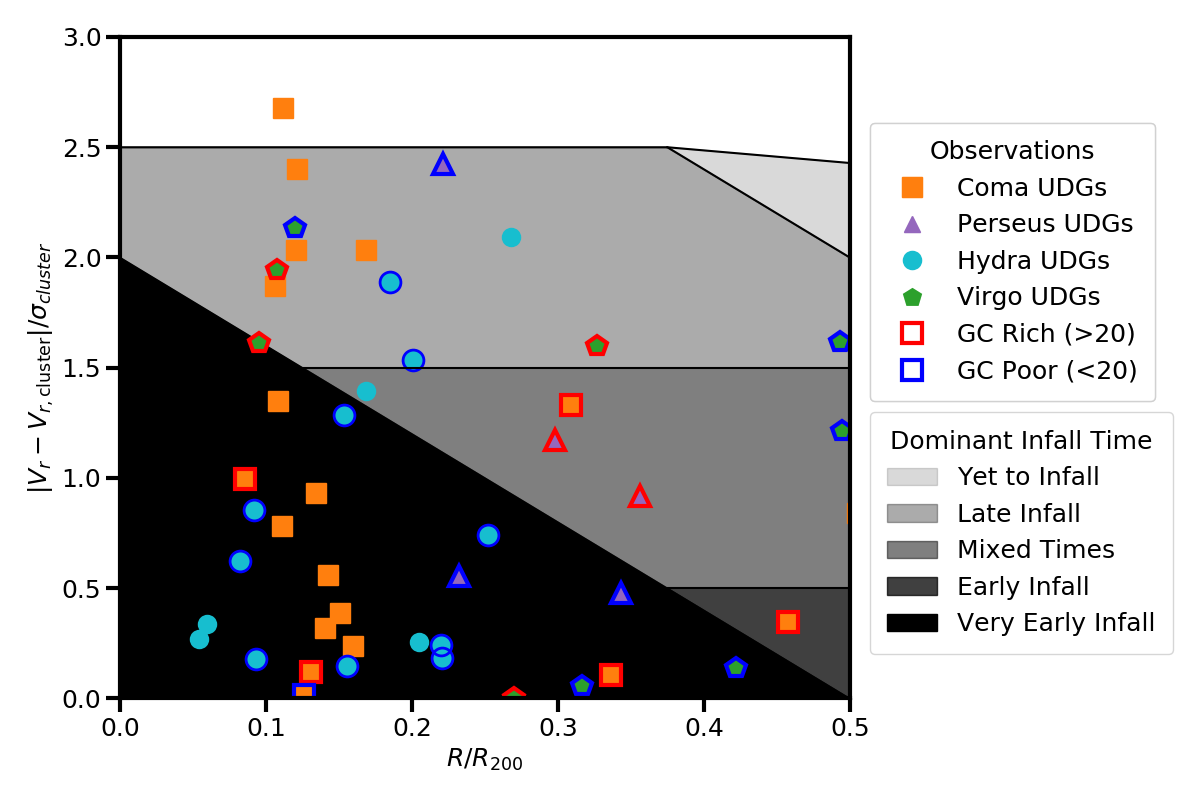}
	\caption{Infall diagnostic diagram for only UDGs in the Hydra~I, Coma, Virgo and Perseus clusters.  Regions of the diagram are shaded  according to their infall times from the cosmological simulations of Rhee et al. (2017).
  As per the legend, UDGs in different clusters are denoted by different symbols. Symbols are outlined in red (if GC-rich) or blue (if GC-poor), and without an outline if the GC properties are unknown. 
 See main text for discussion of selection effects in the UDG samples. Globular cluster (GC) rich UDGs are {\it not} predominately found in the very early infall region, indeed the data suggest that very early infall UDGs tend to be GC-poor.
 }
\end{figure*}

\section{Infall Diagnostic Diagram for UDGs in Several Clusters} 

In Fig. 2 we show the UDGs from the Hydra~I cluster along with those from the literature and coded by globular cluster (GC) richness. Total GC counts for the Hydra~I UDGs are determined in 
\citep{2020A&A...642A..48I, 2022A&A...665A.105L} 
and listed again in Paper I (Iodice et al. 2023, submitted). Literature data comes from \cite{2022MNRAS.510..946G} and the recent work of \cite{2023arXiv230506369T}. The GC counts are almost exclusively based on imaging (i.e. lacking radial velocities) and we
follow \citet{2022MNRAS.510..946G} assigning a somewhat arbitrary separation between rich and poor GC systems at 20 GCs. This corresponds to a halo mass of 10$^{11}$ M$_{\odot}$ using the scaling relation of \cite{2020AJ....159...56B}. Below 20 GCs the scaling relation is less predictive of halo mass due to increased scatter. 
By this definition, all of the UDGs in the Hydra~I cluster are GC-poor (ranging from no GCs for several UDGs to 15 GCs for UDG3) and this is unlikely to change significantly when the full set of MUSE spectroscopic data is available. Given the relatively small stellar mass range of the Hydra~I UDGs, a fixed GC number corresponds closely to a GC system total mass per host galaxy stellar mass. If we assume the same average mass for a GC of 10$^5$ M$_{\odot}$, this ratio is $<$1.2\% for all of the observed Hydra~I UDGs. While 
some Coma cluster UDGs also have a ratio $<$1.2\% the majority have much higher ratios, with up to $\sim$10\% of the galaxy stellar mass in their GC system, see figure 4 of 
\cite{2020MNRAS.492.4874F}.


Before interpreting Fig. 2 there are various caveats and selection effects that should be born in mind. Firstly, we note that some of the literature UDGs lack firm GC counts and their rich/poor status is on the basis of a visual estimate only \citep{2022MNRAS.510..946G}. 
Secondly, the literature sample is subject to sample selection effects. The Coma cluster sample of UDGs comes from studies that have focused on GC-rich galaxies or they have focused on a narrow range in clustercentric radius (i.e. around 0.12~R/R$_{200}$ in the Coma cluster). Observations of the Perseus cluster UDGs have so far avoided the cluster inner  regions. The Virgo UDG sample is relatively small and mostly GC-poor. 
In terms of a selection bias, the Hydra~I UDGs are the closest to being a representative sample of UDGs in the cluster, however only the inner 0.3~R/R$_{200}$ was imaged in \cite{2020A&A...642A..48I}. Thus, we may be missing the late infalling UDGs. We note that 
\cite{2022A&A...665A.105L} estimated a total UDG population out to the virial radius of 48 $\pm$ 10 and so many outer region UDGs, which may be late infallers, remain to be studied. 

The UDG infall diagram does {\it not}  clearly show GC-rich UDGs to be located in earlier infall zones as might be expected in the standard picture of dwarf galaxy quenching due to infall which leads to richer GC systems (as described in the Introduction). Indeed, the opposite trend may be present, such that in the very early infall region  there are 13 GC-poor UDGs and 5 GC-rich ones, whereas outside of this region (but within 0.5 R/R$_{200}$) there are only 6 GC-poor and 6 GC-rich UDGs. Again, we caution that selection and projection effects make conclusions tentative.  


\section{Discussion}

\cite{2018MNRAS.479.3308A} used the phase-space diagram to investigate the infall epoch of UDGs, classical dwarfs and other galaxies in the Coma cluster (a massive, dynamically relaxed cluster). Similar to the Hydra~I cluster, they saw little difference between classical dwarfs and the giant galaxies. For the UDGs, they identified both early and late infallers. A similar situation might be present for Hydra~I UDGs if outer region UDGs were probed. \cite{2018MNRAS.479.3308A} did not include GC richness in their study.

Given the lack of a clear signal for `infall bias' in the GC richness of UDGs alternatives should be further investigated. As noted in the Introduction, quenching at very early times prior to cluster infall should be 
considered. For such UDGs, we would expect very old ages, low metallicities (similar to the metal-poor subpopulation of GCs) and high alpha overabundances (indicative of rapid star formation). A high fraction of mass in GCs relative to field stars might also be expected.
A UDG in the NGC 5846 group,  (NGC5846$\_$UDG1) discovered in VEGAS imaging \citep{2019A&A...626A..66F}, 
may be an example of such a failed galaxy having a remarkable 13\% of its current stellar mass in the form of GCs 
\citep{2022ApJ...927L..28D}.
As noted above, the observed Hydra~I UDGs (from the inner cluster regions) all have less than 1.2\% of their stellar mass in GCs. 

Another possibility is that the 
Hydra~I UDGs are GC-poor because they have been tidally stripped from their host galaxy. This tidal stripping would have to remove most of the dark matter halo before any GCs, since the dark matter is more radially extended than GC systems. Continued stripping would be expected to remove GCs and stars in roughly equal proportions since the radial extent of GC systems for UDGs closely follows that of the galaxy stars. 
As well as operating in clusters, tidal stripping of UDGs may occur in galaxy groups.
We note that UDGs in the field do tend to be GC-poor \citep{2023ApJ...942L...5J}
however  this is unlikely to be due to tidal effects and rather some internal process. 

The Hydra~I UDGs are generally well-fit by a single Sersic profile however a few show hints of asymmetries that might point to a tidal interaction \citep{2020A&A...642A..48I, 2022A&A...659A..92L}. For the one UDG examined in detail by Iodice et al. (2023, submitted) there is some evidence for an isophotal twist in the MUSE data. This might indicate tidal interaction (or a triaxial potential). Furthermore, a Hydra~I UDG first identified by \cite{2008A&A...486..697M} reveals a clear S-shape indicative of ongoing tidal interaction \citep{2012ApJ...755L..13K}.  
In the case of Coma cluster UDGs, \cite{2017ApJ...851...27M} looked  specifically for signs of tidal features via position angle twists in a stacked sample, finding no evidence for such twists. 

\cite{2020MNRAS.494.1848S}
have simulated UDGs in clusters of similar mass to Hydra~I using Illustris-TNG100. They identify two types of UDGs in clusters, i.e. 
Tidal-UDGs and Born-UDGs
(see also \citealt{2019MNRAS.487.5272J}).
The Tidal-UDGs were originally massive galaxies (up to 10$^{10}$ M$_{\odot}$) that have been tidally stripped of stars and puffed-up by  the cluster. Born-UDGs were formed as UDGs outside of the cluster and more recently entered the cluster.
Thus Tidal-UDGs dominate the inner $\sim$0.5R/R$_{200}$ since they were accreted at early times, while Born-UDGs dominate the outer regions with some only recently falling into the cluster. We remind the reader that we only probe out to 0.3R/R$_{200}$ in Hydra~I. 
The \cite{2020MNRAS.494.1848S} 
model would also predict on average higher metallicities, older ages and lower internal velocity dispersions, for a given stellar mass, for their Tidal-UDG compared to the Born-UDGs. 
These stellar population, kinematic, GC colours and dark matter content predicted for Tidal-UDGs can be tested when the full LEWIS dataset is available.

\section{Conclusions}

As part of the LEWIS project (Iodice et al. 2023, in press) we obtained new VLT/MUSE observations of the radial velocities of 
UDGs in the Hydra~I cluster (at D = 51 Mpc). Here we examine the location of Hydra~I UDGs in infall phase-space diagrams based on simulations of cluster galaxies.
We find all of the observed UDGs (and 3 low surface brightness galaxies) to be associated with the cluster. 
From comparison with the 
\cite{2017ApJ...843..128R}
simulations, we conclude that most giants, classical  dwarfs and UDGs fell into the Hydra~I cluster long ago, with UDGs being among the earliest infallers. 
Projection effects in observations and the statistical nature of the infall  diagnostic diagram limit our ability to determine the true fraction of ancient infallers. 
Nevertheless we might expect UDGs in the Hydra~I cluster to reveal old stellar populations consistent with early quenching.

We also compare Hydra~I UDGs with their counterparts in the Coma, Perseus and Virgo clusters in terms of their GC richness. If very early infall into a cluster is associated with enhanced GC richness (as has been suggested for classical dwarf galaxies) then such a trend is expected. The data from these clusters do {\it not} show a clear  trend of GC richness with earlier infall times, indeed the data suggest the opposite trend. If verified by larger and more complete samples, then UDGs may be quenched by a different mechanism than that thought to operate on classical dwarf galaxies. As more data for UDGs is acquired, trends, or the lack of, may become more apparent in an infall diagnostic diagram. 
A future analysis of star formation histories will give an indication of when quenching occurred for the Hydra~I UDGs. Once the full dataset of the LEWIS project is available we will be able to test other mechanisms, such as pre-infall quenching and/or tidal stripping, and their possible role in shaping UDGs and their globular cluster systems.

\section*{Acknowledgements}

We wish to thank the anonymous referee for their 
comments. 
We thank A. Romanowsky, L. Buzzo, L. Haacke and O. Gerhard for useful suggestions. 
This work is based on visitor mode observations
collected at the European Southern Observatory (ESO) La Silla Paranal Observatory and 
collected at the European Southern Observatory under ESO programmes 099.B-0560(A) and 108.222P. INAF
authors acknowledge financial support for the VST project (P.I.: P. Schipani).
DAF thanks the ARC for support via DP220101863. 
Parts of this research were supported by the Australian Research Council Centre of Excellence for All Sky Astrophysics in 3 Dimensions (ASTRO 3D), through project number CE170100013. MC acknowledges support from the INAF-EDGE program (PI Leslie K. Hunt). J.~F-B  acknowledges support through the RAVET project by the grant PID2019-107427GB-C32 from the Spanish Ministry of Science, Innovation and Universities (MCIU), and through the IAC project TRACES which is partially supported through the state budget and the regional budget of the Consejer\'ia de Econom\'ia, Industria, Comercio y Conocimiento of the Canary Islands Autonomous Community.

\section*{Data Availability}

Raw data is available from the ESO archive. 



\bibliographystyle{mnras}
\bibliography{hydra}{}

\bsp	
\label{lastpage}
\end{document}